\documentclass{article}
\usepackage{amsmath, amssymb}
\usepackage{amsfonts}
\usepackage{appendix}
\usepackage{graphicx}
\usepackage{color}
\usepackage{url}
\usepackage{bm}
\usepackage[all]{xy}
\usepackage{arydshln}
\usepackage{enumerate}
\usepackage{hyperref}

\newtheorem{fact}{Fact}
\newtheorem{theorem}{Theorem}
\newtheorem{remark}{Remark} 

\newenvironment{proof}{{\noindent\it Proof}\quad}{\hfill $\square$\par}

\DeclareMathOperator{\W}{W}
\newcommand{\hw}{\text{w}_\text{H}}

\newcommand{\ls}[1]
    {\dimen0=\fontdimen6\the\font\lineskip=#1\dimen0
     \advance\lineskip.5\fontdimen5\the\font
     \advance\lineskip-\dimen0
     \lineskiplimit=0.9\lineskip
     \baselineskip=\lineskip
     \advance\baselineskip\dimen0
     \normallineskip\lineskip\normallineskiplimit\lineskiplimit
     \normalbaselineskip\baselineskip
     \ignorespaces}

\begin{document}

\bibliographystyle{abbrv}

\title{A Note on ``Constructing Bent Functions Outside the Maiorana-McFarland Class Using a General Form of Rothaus"}

\author{Fei Guo$^1$, 
	Zilong Wang$^2$, 
	Guang Gong$^3$\\ 
	\small $^1$  State Key Laboratory of Mathematical Engineering and Advanced Computing, \\[-0.8ex]  
	\small Information Engineering University,  Zhengzhou, 450001, China \\ 
	\small $^2$ State Key Laboratory of Integrated Service Networks, Xidian University, \\[-0.8ex]
	\small Xi'an, 710071, China\\
	\small $^3$Department of Electrical and Computer Engineering, University of Waterloo, \\[-0.8ex]
	\small Waterloo, Ontario N2L 3G1, Canada  \\
	\small\tt guofzz@sina.com, zlwang@xidian.edu.cn, ggong@uwaterloo.ca\\ 
}

\maketitle

\thispagestyle{plain} \setcounter{page}{1}

\ls{1.5}

\begin{abstract} 
	In 2017, Zhang et al. proposed a question (not open problem) and two open problems in [IEEE TIT 63 (8): 5336--5349, 2017] about constructing bent functions by using Rothaus' construction. 
	In this note, we prove that the sufficient conditions of Rothaus' construction are also necessary, which answers their question. 
	Besides, we demonstrate that the second open problem, which considers the iterative method of constructing bent functions by using Rothaus' construction, has only a trivial solution. 
	It indicates that all bent functions obtained by using Rothaus' construction iteratively can be generated from the direct sum of an initial bent function and a quadratic bent function. 
	This directly means that Zhang et al.'s construction idea 
	makes no contribution to the construction of bent functions. 
	To compensate the weakness of their work, we propose an iterative construction of bent functions by using a secondary construction in [DCC 88: 2007--2035, 2020]. 
\end{abstract} 

{\bf Index Terms } 
Bent function, Rothaus' construction, Walsh-Hadamard transform, Boolean function. 

\section{Introduction} 
In this note, addition over the real number field is denoted by $+$ while addition over the finite field $\mathbb{F}_2$ is denoted by $\oplus$. 
Let $\mathbb{F}_2^n$ denote the $n$-dimensional vector space of                                                                                                                                                                                                                                                                                                                                                                                                                                                                                                                                                                                                                                                                                                                                                                                                                                                                                                                                                                                                                                                                                                                                                                                                                                                                                                                                                                                                                                                                                                                                                                                                                                                                                                                                                                                                                                                                                                                                                                                                                                                                                                                                                                                                                                                                                                                                                                                                                                                                                                                                                                                                                                                                                                                                                                                                                                                                                                                                                                                                                                                                                                                                                                                                                   $\mathbb{F}_2$, where $n$ is a positive integer. 
A Boolean function on $n$ variables is a mapping from $\mathbb{F}_2^n$ to $\mathbb{F}_2$, which can be uniquely represented as the multivariate polynomial: 
\[
f(x_1,\cdots,x_n) = \bigoplus_{\bm{\mathrm{\alpha}}=(a_1,\cdots,a_n)\in\mathbb{F}_2^{n}}\lambda_{\bm{\mathrm{\alpha}}} \prod_{i=1}^n x_i^{a_i}, 
\]
where $\lambda_{\bm{\mathrm{\alpha}}} \in \mathbb{F}_2$. 
The Walsh-Hadamard transform of $f$ at $\bm{\mathrm{u}}= (u_1, \cdots, u_n) \in \mathbb{F}_2^{n}$ is defined by 
\[
\W_{f}(\bm{\mathrm{u}})=\sum_{\bm{\mathrm{x}} = (x_1, \cdots, x_n) \in{\mathbb{F}_2^{n}}}(-1)^{f(\bm{\mathrm{x}})\oplus \bm{\mathrm{u}} \cdot \bm{\mathrm{x}}}, 
\]
where $\bm{\mathrm{u}} \cdot \bm{\mathrm{x}} = \bigoplus_{i=1}^n u_ix_i$ is the usual inner product of $\bm{\mathrm{u}}$ and $\bm{\mathrm{x}}$ over $\mathbb{F}_2$. 

A Boolean function $f$ on $n$ variables ($n$ even) is called bent if $|\W_f(\bm{\mathrm{u}})| = 2^{\frac{n}{2}}$ for all $\bm{\mathrm{u}} \in \mathbb{F}_2^n$. 
Bent functions play an important role in cryptography, coding theory and design of sequences. 
Next, we review Rothaus' secondary construction of bent functions. 
\begin{fact} \rm \rm (\cite{Rothaus1976}) 
	\label{fact:RothausConstruction} 
	Let $n$ be a positive even integer, $\bm{\mathrm{x}} \in \mathbb{F}_2^n$, and $x_{n+1}, x_{n+2} \in \mathbb{F}_2$. 
	Let $A, B, C$ be bent functions on $\mathbb{F}_2^n$ such that $A\oplus B \oplus C$ is also bent. 
	Then $f: \mathbb{F}_2^n \times \mathbb{F}_2 \times \mathbb{F}_2 \rightarrow \mathbb{F}_2$ defined by 
	\begin{equation} \label{equation:RothausConstruction} 
	f(\bm{\mathrm{x}}, x_{n+1}, x_{n+2}) = A(\bm{\mathrm{x}})B(\bm{\mathrm{x}})\oplus A(\bm{\mathrm{x}})C(\bm{\mathrm{x}})\oplus B(\bm{\mathrm{x}})C(\bm{\mathrm{x}})\oplus x_{n+1}x_{n+2} \oplus \left[A(\bm{\mathrm{x}})\oplus B(\bm{\mathrm{x}})\right]x_{n+1}\oplus \left[A(\bm{\mathrm{x}})\oplus C(\bm{\mathrm{x}})\right]x_{n+2} 
	\end{equation}
	is a bent function. 
\end{fact}

In Section III of \cite{Zhang2017}, 
Zhang et al. investigated a special subclass of Rothaus' construction, in which two bent functions of the three that differ by a characteristic function of a suitably chosen $\frac{n}{2}$-dimensional subspace. 
In this special class, the given conditions are proved to be both necessary and sufficient. 
Then, they proposed a question in \cite[Page 5339]{Zhang2017}: 
{\em 
	Whether the conditions that the initial functions and their sum are bent are in general also necessary in Rothaus' construction. 
} 
They further showed that the special subclass of Rothaus' construction is efficient to generate bent functions outside the completed Maiorana-McFarland class by specifying initial functions in the class $\mathcal{D}$. 
They also analyzed the conditions of the produced bent functions from the method of Rothaus not belonging to the completed Maiorana-McFarland class when using initial functions in the class $\mathcal{C}$. 

In Section IV of \cite{Zhang2017}, Zhang et al. proposed an idea of constructing an infinite sequence of bent functions by using Rothaus' construction iteratively. 
For completeness, we restate their construction process here. 
Let $\bm{\mathrm{x}} \in \mathbb{F}_2^n$, and $x_{n+1}, x_{n+2} \in \mathbb{F}_2$. 
Define two other bent functions on $\mathbb{F}_2^{n+2}$ similarly to $f$ in (\ref{equation:RothausConstruction}): 
\[
f'(\bm{\mathrm{x}}, x_{n+1}, x_{n+2}) = A(\bm{\mathrm{x}})B(\bm{\mathrm{x}})\oplus A(\bm{\mathrm{x}})C(\bm{\mathrm{x}})\oplus B(\bm{\mathrm{x}})C(\bm{\mathrm{x}})\oplus x_{n+1}x_{n+2} \oplus \left[B(\bm{\mathrm{x}})\oplus C(\bm{\mathrm{x}})\right]x_{n+1}\oplus \left[A(\bm{\mathrm{x}})\oplus B(\bm{\mathrm{x}})\right]x_{n+2}, 
\]
\begin{equation} \label{equation:Rothaus_f'f''} 
f''(\bm{\mathrm{x}}, x_{n+1}, x_{n+2}) = A(\bm{\mathrm{x}})B(\bm{\mathrm{x}})\oplus A(\bm{\mathrm{x}})C(\bm{\mathrm{x}})\oplus B(\bm{\mathrm{x}})C(\bm{\mathrm{x}})\oplus x_{n+1}x_{n+2} \oplus \left[A(\bm{\mathrm{x}})\oplus C(\bm{\mathrm{x}})\right]x_{n+1}\oplus \left[B(\bm{\mathrm{x}})\oplus C(\bm{\mathrm{x}})\right]x_{n+2}. 
\end{equation} 
Then, one knows $f, f'$ and $f''$ are bent functions if $A, B, C$ and $A\oplus B\oplus C$ are all bent functions. 
Further, using $f, f'$ and $f''$ as initial functions, 
then an $(n+4)$-variable bent function can be generated from Rothaus' construction if $f\oplus f'\oplus f''$ is bent again. 
It is easily verified that 
\[
(f\oplus f'\oplus f'')(\bm{\mathrm{x}}, x_{n+1}, x_{n+2}) = A(\bm{\mathrm{x}})B(\bm{\mathrm{x}})\oplus A(\bm{\mathrm{x}})C(\bm{\mathrm{x}})\oplus B(\bm{\mathrm{x}})C(\bm{\mathrm{x}})\oplus x_{n+1}x_{n+2}. 
\]
Due to the special structure of this expression, we know $f\oplus f'\oplus f''$ is bent if and only if $AB\oplus AC\oplus BC$ is bent. 
Some examples of such triple bent functions $A, B$ and $C$ are presented in \cite{Mesnager2014} such that $AB\oplus AC\oplus BC$ is bent. 
Consequently, with these $A, B$ and $C$, $f, f'$ and $f''$ obtained from (\ref{equation:RothausConstruction}) and (\ref{equation:Rothaus_f'f''}) can be used as initial functions of Rothaus' construction again.  

Let $\bm{\mathrm{X}} = (\bm{\mathrm{x}}, x_{n+1}, x_{n+2})$ and $x_{n+3}, x_{n+4} \in \mathbb{F}_2$. 
Using $f, f'$ and $f''$ as initial functions, the following three bent functions on $(n+4)$ variables are obtained from Rothaus' construction:  
\[
\begin{split}
g(\bm{\mathrm{X}}, x_{n+3}, x_{n+4}) &= (ff'\oplus ff''\oplus f'f'')(\bm{\mathrm{X}})\oplus x_{n+3}x_{n+4}\oplus [f(\bm{\mathrm{X}})\oplus f'(\bm{\mathrm{X}})]x_{n+3} \oplus [f(\bm{\mathrm{X}})\oplus f''(\bm{\mathrm{X}})]x_{n+4}, \\ 
g'(\bm{\mathrm{X}}, x_{n+3}, x_{n+4}) &= (ff'\oplus ff''\oplus f'f'')(\bm{\mathrm{X}})\oplus x_{n+3}x_{n+4}\oplus [f'(\bm{\mathrm{X}})\oplus f''(\bm{\mathrm{X}})]x_{n+3} \oplus [f(\bm{\mathrm{X}})\oplus f'(\bm{\mathrm{X}})]x_{n+4}, \\ 
g''(\bm{\mathrm{X}}, x_{n+3}, x_{n+4}) &= (ff'\oplus ff''\oplus f'f'')(\bm{\mathrm{X}})\oplus x_{n+3}x_{n+4}\oplus [f(\bm{\mathrm{X}})\oplus f''(\bm{\mathrm{X}})]x_{n+3} \oplus [f'(\bm{\mathrm{X}})\oplus f''(\bm{\mathrm{X}})]x_{n+4}. 
\end{split}
\]
Once again, using $g, g'$ and $g''$ as initial functions, 
an $(n+6)$-variable bent function can be produced from Rothaus' construction if $g\oplus g'\oplus g''$ is bent again. 
We have 
\[
(g\oplus g' \oplus g'')(\bm{\mathrm{X}}, x_{n+3}, x_{n+4}) = f(\bm{\mathrm{X}})f'(\bm{\mathrm{X}}) \oplus f(\bm{\mathrm{X}})f''(\bm{\mathrm{X}})\oplus f'(\bm{\mathrm{X}})f''(\bm{\mathrm{X}}) \oplus x_{n+3}x_{n+4}, 
\]
which is bent if and only if $ff' \oplus ff''\oplus f'f''$ is bent. 
It is easily verified that 
\[
(ff' \oplus ff''\oplus f'f'')(\bm{\mathrm{X}}) = D(\bm{\mathrm{x}})\oplus x_{n+1}x_{n+2} \oplus (x_{n+1}\oplus x_{n+2} \oplus x_{n+1}x_{n+2})(D\oplus A\oplus B \oplus C)(\bm{\mathrm{x}}), 
\]
where $D(\bm{\mathrm{x}}) = A(\bm{\mathrm{x}})B(\bm{\mathrm{x}})\oplus A(\bm{\mathrm{x}})C(\bm{\mathrm{x}})\oplus B(\bm{\mathrm{x}})C(\bm{\mathrm{x}})$. 
Clearly, $ff' \oplus ff''\oplus f'f''$ can be expressed as the concatenation of four functions: $D,\ A\oplus B\oplus C,\ A\oplus B\oplus C$ and $A\oplus B\oplus C\oplus 1$. 
Hence, we know $ff' \oplus ff''\oplus f'f''$ is bent if and only if $D = A\oplus B\oplus C$, i.e., $AB\oplus AC\oplus BC = A\oplus B \oplus C$. 
This is trivially satisfied for $A = B = C$, and finding non-trivial three initial functions satisfying this condition seems to be hard work. 
Zhang et al. put forward an open problem \cite[Open Problem 2]{Zhang2017}: 
{\em 
	It would be of interest to specify conditions on initial functions $A, B, C$ along with suitably defined $f$, $f'$ and $f''$, where $f'$ and $f''$ are symmetric versions of $f$, that would give rise to an infinite sequence of bent functions stemming from the method of Rothaus. 
}

In this note, first, 
we prove that the sufficient conditions of Rothaus' construction are also necessary, which answers Zhang et al.'s question in \cite[Page 5339]{Zhang2017}. 
Second, we show that the open problem in \cite[Open Problem 2]{Zhang2017} only has a trivial solution, i.e., $AB\oplus AC\oplus BC = A\oplus B\oplus C$ if and only if $A = B = C$. 
However, the trivial solution makes no contribution to the construction of bent functions. 
Third, we propose an iterative construction of bent functions by using the secondary construction in \cite[Theorem 4.3]{Hodzic2020}. 

\section{Main Results}

To answer Zhang et al.'s question and open problem, we need the following preparatory result on the Walsh-Hadamard transform of the composition of Boolean functions. 
\begin{fact} \rm (\cite[Theorem 3]{Gupta2005}) 
	\label{fact:Composition_Walsh}
	Let $h(z_1, \cdots, z_k)$ be a $k$-variable Boolean function, and $F(\bm{\mathrm{x}})=(f_1(\bm{\mathrm{x}}), f_2(\bm{\mathrm{x}}),\cdots, f_k(\bm{\mathrm{x}}))$ a multi-output function from $\mathbb{F}_2^n$ to $\mathbb{F}_2^k$, where $f_i$'s $(1\le i\le k)$ are Boolean functions on $n$ variables.
	Then the Walsh-Hadamard transform of the composition function $(h\circ F)(\bm{\mathrm{x}})$ at $\bm{\mathrm{u}} \in \mathbb{F}_2^{n}$ is given by
	\[
	\W_{(h\circ F)} (\bm{\mathrm{u}}) = \frac{1}{2^k} \sum_{\bm{\mathrm{\omega}} \in {\mathbb{F}_2^{k}}} \W_{h}(\bm{\mathrm{\omega}}) \W_{\bm{\mathrm{\omega}} \cdot (f_1, f_2,\cdots, f_k)}(\bm{\mathrm{u}}). 
	\] 
\end{fact}

First, we answer Zhang et al.'s question in \cite[Page 5339]{Zhang2017}. 

\begin{theorem} \rm \label{theorem:RothausConstruction} 
	Let $n$ be a positive even integer, $\bm{\mathrm{x}} \in \mathbb{F}_2^n$, and $x_{n+1}, x_{n+2} \in \mathbb{F}_2$. 
	Let $A, B, C$ be Boolean functions on $\mathbb{F}_2^n$. 
	Then the $(n+2)$-variable function $f$ defined in (\ref{equation:RothausConstruction}) 
	is bent if and only if $A, B, C$ and $A\oplus B\oplus C$ are all bent. 
\end{theorem}

\begin{proof}
	The sufficiency holds by Fact \ref{fact:RothausConstruction}. 
	Then we prove the necessity. 
	Let $h$ be a $5$-variable function defined by 
	\[
	h(x_1, x_2, x_3, x_4, x_5) = x_1x_2\oplus x_1x_3\oplus x_2x_3 \oplus x_4x_5\oplus (x_1\oplus x_2)x_4\oplus (x_1\oplus x_3)x_5. 
	\] 
	Obviously, $f$ in (\ref{equation:RothausConstruction}) can be expressed by $f(\bm{\mathrm{x}}, x_{n+1}, x_{n+2})=h\circ (A(\bm{\mathrm{x}}), B(\bm{\mathrm{x}}), C(\bm{\mathrm{x}}), x_{n+1}, x_{n+2})$. 
	By a direct calculation, we obtain
	\[
	\W_h(1,0,0,0,0)=16,\ \W_h(0,1,0,0,1)=16,\ \W_h(0,0,1,1,0)=16,\ \W_h(1,1,1,1,1)=-16, 
	\] 
	and the Walsh-Hadamard transform of $h$ vanishes at other positions. 
	By Fact \ref{fact:Composition_Walsh}, the Walsh-Hadamard transform of $f$ at $(\bm{\mathrm{u}}, u_{n+1}, u_{n+2}) \in \mathbb{F}_2^n \times \mathbb{F}_2 \times \mathbb{F}_2$ is given by  
	\begin{align}
	\W_f(\bm{\mathrm{u}}, u_{n+1}, u_{n+2}) =& \frac{1}{2^5} \sum_{\bm{\mathrm{\omega}} \in {\mathbb{F}_2^{5}}} \W_{h}(\bm{\mathrm{\omega}}) \W_{\bm{\mathrm{\omega}} \cdot (A, B, C, x_{n+1}, x_{n+2})}(\bm{\mathrm{u}}, u_{n+1}, u_{n+2}) \notag \\ 
	=& \frac{1}{2}\left[\W_{A(\bm{\mathrm{x}})}(\bm{\mathrm{u}}, u_{n+1}, u_{n+2}) + \W_{B(\bm{\mathrm{x}}) \oplus x_{n+2}}(\bm{\mathrm{u}}, u_{n+1}, u_{n+2}) + \right. \notag \\
	& \hspace{0.4cm}\left. \W_{C(\bm{\mathrm{x}}) \oplus x_{n+1}}(\bm{\mathrm{u}}, u_{n+1}, u_{n+2}) + \W_{(A\oplus B\oplus C)(\bm{\mathrm{x}}) \oplus x_{n+1}\oplus x_{n+2}}(\bm{\mathrm{u}}, u_{n+1}, u_{n+2}) \right] \notag \\
	=& \begin{cases}
	2\W_A(\bm{\mathrm{u}}), \hspace{1.2cm} (u_{n+1}, u_{n+2})=(0,0); \\ 
	2\W_B(\bm{\mathrm{u}}), \hspace{1.2cm} (u_{n+1}, u_{n+2})=(0,1); \\
	2\W_C(\bm{\mathrm{u}}), \hspace{1.2cm} (u_{n+1}, u_{n+2})=(1,0); \\
	-2\W_{A\oplus B\oplus C}(\bm{\mathrm{u}}),\ (u_{n+1}, u_{n+2})=(1,1). \\
	\end{cases} \label{equation:WalshSpectrum_f}
	\end{align}
	Assume $f$ is bent, i.e., $|\W_f(\bm{\mathrm{u}}, u_{n+1}, u_{n+2})| = 2^{\frac{n}{2}+1}$ for all $(\bm{\mathrm{u}}, u_{n+1}, u_{n+2}) \in \mathbb{F}_2^n \times \mathbb{F}_2 \times \mathbb{F}_2$. 
	By (\ref{equation:WalshSpectrum_f}), we know all of $|\W_A(\bm{\mathrm{u}})|, |\W_B(\bm{\mathrm{u}})|, |\W_C(\bm{\mathrm{u}})|$ and $|\W_{A\oplus B\oplus C}(\bm{\mathrm{u}})|$ are equal to $2^{\frac{n}{2}}$ for all $\bm{\mathrm{u}}\in \mathbb{F}_2^n$. 
	That is to say, $A, B, C$ and $A\oplus B\oplus C$ are all bent functions. 
	Thus, the necessity holds. 
\end{proof}

\begin{remark} \rm 
	Theorem \ref{theorem:RothausConstruction} shows that the conditions of Rothaus' construction are both necessary and sufficient. 
	This answers Zhang et al.'s question in \cite[Page 5339]{Zhang2017}. 
\end{remark}

Next, we prove that the open problem in \cite[Open Problem 2]{Zhang2017} has only a trivial solution. 

\begin{theorem} \rm \label{theorem:maintheorem}
	Let $n$ be a positive integer. 
	Three $n$-variable Boolean functions $A, B$ and $C$ satisfy 
	\begin{equation*}
	AB\oplus AC\oplus BC = A\oplus B\oplus C 
	\end{equation*} 
	if and only if $A = B = C$. 
\end{theorem}

\begin{proof} 
	The sufficiency can be easily verified. 
	In what follows, we prove the necessity. 
	Let $h$ be a function on three variables defined as
	\[
	h(x_1, x_2, x_3)=x_1x_2\oplus x_1x_3\oplus x_2x_3\oplus x_1\oplus x_2\oplus x_3. 
	\]
	By a direct calculation, we obtain 
	\[
	\W_h(0,0,0) = -4, \ \W_h(0,1,1) = 4, \ \W_h(1,0,1) = 4, \ \W_h(1,1,0) = 4, 
	\]
	and the Walsh-Hadamard transform of $h$ vanishes at other positions. 	
	Let $\bm{\mathrm{x}} \in \mathbb{F}_2^n$. 
	Define a Boolean function $\mathfrak{f}$ on $\mathbb{F}_2^n$ as  
	\[
	\mathfrak{f}(\bm{\mathrm{x}})=h\circ (A(\bm{\mathrm{x}}), B(\bm{\mathrm{x}}), C(\bm{\mathrm{x}})) = A(\bm{\mathrm{x}})B(\bm{\mathrm{x}})\oplus A(\bm{\mathrm{x}})C(\bm{\mathrm{x}})\oplus B(\bm{\mathrm{x}})C(\bm{\mathrm{x}})\oplus A(\bm{\mathrm{x}})\oplus B(\bm{\mathrm{x}})\oplus C(\bm{\mathrm{x}}). 
	\]
	By Fact \ref{fact:Composition_Walsh}, the Walsh-Hadamard transform of $\mathfrak{f}$ at $\bm{\mathrm{u}}\in \mathbb{F}_2^n$ is given by  
	\begin{align}
	\W_{\mathfrak{f}}(\bm{\mathrm{u}}) =& \frac{1}{2^3} \sum_{\bm{\mathrm{\omega}} \in {\mathbb{F}_2^{3}}} \W_{h}(\bm{\mathrm{\omega}}) \W_{\bm{\mathrm{\omega}} \cdot (A, B, C)}(\bm{\mathrm{u}}) \notag \\ 
	=& \frac{1}{2}\left[-\W_{0}(\bm{\mathrm{u}}) + \W_{A\oplus B}(\bm{\mathrm{u}})+ \W_{A\oplus C}(\bm{\mathrm{u}})+ \W_{B\oplus C}(\bm{\mathrm{u}})\right]. \label{equation:WalshSpectrum_f2}
	\end{align}
	Assume $AB\oplus AC\oplus BC = A\oplus B\oplus C$, i.e., $\mathfrak{f}$ is the constant zero function. 
	Then we have $\W_{\mathfrak{f}}(\bm{0}_n) = 2^n$, where $\bm{0}_n$ is the all-zero vector in $\mathbb{F}_2^n$. 
	Thus, from (\ref{equation:WalshSpectrum_f2}) we have  
	\[
	 \frac{1}{2}\left[-\W_{0}(\bm{0}_n) + \W_{A\oplus B}(\bm{0}_n)+ \W_{A\oplus C}(\bm{0}_n)+ \W_{B\oplus C}(\bm{0}_n)\right] = 2^n.  
	\]
	That is, 
	\begin{equation} 
	\label{equation:Intermediate}
	\W_{A\oplus B}(\bm{0}_n)+ \W_{A\oplus C}(\bm{0}_n)+ \W_{B\oplus C}(\bm{0}_n) = 3 \cdot 2^n.  
	\end{equation}
	
	On the other hand, it is easily verified that $\W_{g}(\bm{0}_n) + 2\hw(g) = 2^n$ holds 
	for arbitrary Boolean function $g$ on $\mathbb{F}_2^n$, where $\hw(g) = |\{\bm{\mathrm{x}} \in \mathbb{F}_2^n : g(\bm{\mathrm{x}}) = 1\}|$ is the Hamming weight of $g$. 
	Then, (\ref{equation:Intermediate}) is equivalent to   
	\[
	\hw(A\oplus B)+ \hw(A\oplus C)+ \hw(B\oplus C) = 0, 
	\]
	which immediately indicates $A=B=C$. 
	Thus, the necessity holds. 
\end{proof}

\begin{remark} \rm 
	Theorem \ref{theorem:maintheorem} shows that $A=B=C$ is the only solution to $AB\oplus AC\oplus BC = A\oplus B\oplus C$. 
	In this case, $f$ in (\ref{equation:RothausConstruction}) becomes $f(\bm{\mathrm{x}}, x_{n+1}, x_{n+2}) = A(\bm{\mathrm{x}}) \oplus x_{n+1}x_{n+2}$, which is actually the direct sum of two bent functions $A(\bm{\mathrm{x}})$ and $x_{n+1}x_{n+2}$. 
	Moreover, it is obvious that $f'' = f' = f$. 
	It indicates that bent functions produced from Rothaus' construction iteratively are always the direct sum of two bent functions.
	Therefore, the iterative construction method of Zhang et al. makes no contribution to the construction of bent functions. 
\end{remark}

Finally, we propose an iterative method for constructing bent functions. 
In \cite{Hodzic2020}, Hod$\check{\text{z}}$i$\acute{\text{c}}$ et al. proposed a secondary construction of bent functions which employs the same conditions as those of Rothaus' construction. 
The difference between Rothaus' construction and Hod$\check{\text{z}}$i$\acute{\text{c}}$ et al.'s construction  is that the former produces bent functions on $(n+2)$ variables while the latter produces bent functions on $(n+4)$ variables. 

\begin{fact} \rm (\cite[Theorem 4.3]{Hodzic2020})
	\label{fact:generalizedRothaus}
	Let $n$ be a positive even integer, $\bm{\mathrm{x}} \in \mathbb{F}_2^n$, and $x_{n+1}, x_{n+2}, x_{n+3}, x_{n+4} \in \mathbb{F}_2$. 
	Let $A, B, C$ be bent functions on $\mathbb{F}_2^n$ such that $A\oplus B\oplus C$ is also bent. 
	Then $g: \mathbb{F}_2^n \times \mathbb{F}_2 \times \mathbb{F}_2  \times \mathbb{F}_2 \times \mathbb{F}_2 \rightarrow \mathbb{F}_2$ defined by 
	\begin{equation} \label{equation:generalizedRothaus}
	g(\bm{\mathrm{x}}, x_{n+1}, \cdots, x_{n+4}) = B(\bm{\mathrm{x}})(x_{n+1}\oplus x_{n+2})\oplus A(\bm{\mathrm{x}})(1\oplus x_{n+1}\oplus x_{n+3})\oplus (C(\bm{\mathrm{x}})\oplus x_{n+1})(x_{n+2}\oplus x_{n+3})\oplus (x_{n+1}\oplus x_{n+2})x_{n+4}
	\end{equation}
	is a bent function. 
\end{fact}

We define the other two bent functions on $\mathbb{F}_2^{n+4}$ similarly to $g$: 
\[
g'(\bm{\mathrm{x}}, x_{n+1}, \cdots, x_{n+4}) = A(\bm{\mathrm{x}})(x_{n+1}\oplus x_{n+2})\oplus C(\bm{\mathrm{x}})(1\oplus x_{n+1}\oplus x_{n+3})\oplus (B(\bm{\mathrm{x}})\oplus x_{n+1})(x_{n+2}\oplus x_{n+3})\oplus (x_{n+1}\oplus x_{n+2})x_{n+4}, 
\]
\begin{equation} \label{equation:generalizedRothaus_g'g''}
g''(\bm{\mathrm{x}}, x_{n+1}, \cdots, x_{n+4}) = C(\bm{\mathrm{x}})(x_{n+1}\oplus x_{n+2})\oplus B(\bm{\mathrm{x}})(1\oplus x_{n+1}\oplus x_{n+3})\oplus (A(\bm{\mathrm{x}})\oplus x_{n+1})(x_{n+2}\oplus x_{n+3})\oplus (x_{n+1}\oplus x_{n+2})x_{n+4}. 
\end{equation}
It is readily verified that 
\[
(g\oplus g'\oplus g'')(\bm{\mathrm{x}}, x_{n+1}, \cdots, x_{n+4})=A(\bm{\mathrm{x}})\oplus B(\bm{\mathrm{x}})\oplus C(\bm{\mathrm{x}})\oplus x_{n+1}x_{n+4}\oplus (x_{x+1}\oplus x_{x+4}, x_{n+1})\cdot (x_{n+2}, x_{n+3}), 
\]
which is a bent function obviously since $x_{n+1}x_{n+4}\oplus (x_{x+1}\oplus x_{x+4}, x_{n+1})\cdot (x_{n+2}, x_{n+3})$ is a bent function in the Maiorana-McFarland class over $\mathbb{F}_2^4$.  
Thus, using $g, g'$ and $g''$ as initial functions, three bent functions on $(n+8)$ variables can be produced from (\ref{equation:generalizedRothaus}) and (\ref{equation:generalizedRothaus_g'g''}). 
Further, by carrying out Fact \ref{fact:generalizedRothaus} iteratively, i.e., with three previous resulting functions serving as inputs of next operation, an infinite sequence of bent functions on $(n + 4k)$ variables can be generated, where $k=1,2,3, \cdots$. 

\begin{remark} \rm 
	From the above analysis, we know that the iterative method of constructing bent functions works on Hod$\check{\text{z}}$i$\acute{\text{c}}$ et al.'s construction in Fact \ref{fact:generalizedRothaus} instead of Rothaus' construction in Fact \ref{fact:RothausConstruction}. 
\end{remark}


\end{document}